\documentclass[showpacs,aps,prd,preprint,nofootinbib,showkeys,unsortedaddress,raggedbottom]{revtex4-1}
\pdfoutput=1

\usepackage{bm}
\usepackage{amsmath}
\usepackage{graphicx}
\usepackage{subfigure}
\usepackage{float}
\usepackage[usenames,dvipsnames]{color}
\definecolor{darkblue}{RGB}{0,0,196}
\usepackage[colorlinks=true,linkcolor=darkblue,citecolor=darkblue,urlcolor=darkblue]{hyperref}

\usepackage{setspace}
\usepackage{footmisc}
\usepackage[makeroom]{cancel}

\def\be{\begin{equation}}
\def\ee{\end{equation}}
\def\ba{\begin{eqnarray}}
\def\ea{\end{eqnarray}}

\begin{document}

\title{Quark self-energy in an ellipsoidally anisotropic quark-gluon plasma}

\author{Babak S. Kasmaei, Mohammad Nopoush, and Michael Strickland}
\affiliation{Department of Physics, Kent State University, Kent, OH 44242 United States}

\begin{abstract}
We calculate the quark self-energy in a quark-gluon plasma that possesses an ellipsoidal momentum-space anisotropy in the local rest frame. By introducing additional transverse momentum anisotropy parameters into the parton distribution functions, we generalize previous results which were obtained for the case of a spheroidal anisotropy. Our results demonstrate that the presence of anisotropies in the transverse directions affects the real and imaginary parts of quark self-energy and, consequently, the self-energy depends on both the polar and azimuthal angles in the local rest frame of the matter.  Our results for the quark self-energy set the stage for the calculation of the effects of ellipsoidal momentum-space anisotropy on quark-gluon plasma photon spectra and collective flow.
\end{abstract}
\pacs{11.10Wx, 12.38Mh, 25.75.-q}
\maketitle
\newpage

\small

\section{Introduction}

Quantum chromodynamics (QCD) is the principal model for strong interactions and has been successfully applied to address various physical phenomena from high-energy deep inelastic scattering experiments \cite{Buras:1980} to low-energy masses of light hadrons \phantomsection\cite{Fodor:2012}, albeit the latter within a discretized version of the theory on the lattice. Despite these successes, there are still many cases in which the full application of QCD to justify or predict experimental data cannot yet be performed due to mathematical and/or computational difficulties. As a result, the characteristics, and even the existence in some cases, of the various phases of strongly interacting matter are not yet fully understood. The riddle of transitions between these phases is even more challenging.

High-energy heavy-ion collision experiments at the Relativistic Heavy Ion Collider (RHIC) and the Large Hadron Collider (LHC) are being performed in order to improve our understanding of the physics of quarks and gluons subject to extreme conditions. After the nuclear collision and prior to the detection of the produced hadrons, it is commonly believed that a very hot and short-lived quark-gluon plasma (QGP) phase is produced \cite{Gyulassy:2004zy, Yagi-Hatsuda, Munzinger:2007, Jacak-Muller:2012, Heinz:2013th} in which one may use a partonic description to understand the collective behavior of the system. Some of the main features of such a system are viscosity, momentum-anisotropy, and non-equilibrium evolution. Hydrodynamic models of  QGP evolution have been used as phenomenological tools to describe various aspects of the detected spectrum of hadrons, leptons, and photons produced in heavy-ion collision experiments.  In recent years, viscous hydrodynamics \cite{Muronga:2001zk,Muronga:2003ta,Muronga:2004sf,Heinz:2005bw,Baier:2006um,Romatschke:2007mq,Baier:2007ix,
Dusling:2007gi,Luzum:2008cw,Song:2008hj,Heinz:2009xj,Bozek:2009ty,Bozek:2009dw,PeraltaRamos:2009kg,
PeraltaRamos:2010je,Denicol:2010tr,Denicol:2010xn,Schenke:2010rr,Schenke:2011tv,Bozek:2011wa,Bozek:2011ua,
Niemi:2011ix,Niemi:2012ry,Bozek:2012qs,Denicol:2012cn,Denicol:2012es,PeraltaRamos:2012xk,Calzetta:2014hra,
Denicol:2014vaa,Florkowski:2015lra,Ryu:2015vwa,Niemi:2015voa,Niemi:2015bpj} and anisotropic hydrodynamics \cite{Martinez:2010sc,Florkowski:2010cf,Ryblewski:2010bs,Martinez:2010sd,Ryblewski:2011aq,
Florkowski:2011jg,Martinez:2012tu,Ryblewski:2012rr,Florkowski:2012as,Florkowski:2013uqa,Ryblewski:2013jsa,
Bazow:2013ifa,Tinti:2013vba,Florkowski:2014bba,Florkowski:2014txa,Nopoush:2014pfa,Denicol:2014mca,Nopoush:2014pfa,
Tinti:2015xra,Bazow:2015cha,Nopoush:2014qba,Nopoush:2015yga,Bazow:2015zca,Alqahtani:2015qja,Alqahtani:2016rth,
Florkowski:2015cba,Molnar:2016vvu} have provided indirect measures of the QGP's features such as effective temperature, viscosity, and degree of momentum-space anisotropy; however, constraining QGP initial conditions experimentally is difficult if one restricts attention to late-time hadronic production.  Since early-time electromagnetic emissions are not significantly affected during their propagation through the QGP, electromagnetic probes such as photons and dileptons have been considered as ideal probes of the early-time dynamics of the system~\cite{Shuryak:1978ij,Domokos:1980ba,Kajantie:1981wg,Gale:1988ym,Neubert:1989hu,Shuryak:1992bt,Kampfer:1992bb,Kapusta:1992uy,Dumitru:1993vz,Strickland:1994rf,Chatterjee:2005de,Chatterjee:2007xk,Gale2013,Vujanovic:2013jpa,Shen:2013vja,Shen:2014nfa,Ryblewski:2015hea,Shen2016,Vujanovic:2016anq,Bhattacharya:2016}. 

To analyse the properties of a hot QGP, one can use finite temperature field theory \cite{Kapusta-Gale} and, for the treatment of non-equilibrium dynamics, one can use the real-time formalism of quantum field theory~\cite{AshokDas}. The quark self-energy and dispersion relations in a hot isotropic environment have been calculated originally in Ref.~\cite{Weldon:1982}. The imaginary part of the self-energy is related to the generalized decay and inverse decay rates which provide information about the emission or absorption of particles \cite{Weldon:1983,Bodeker:2015exa}. The collective fermionic modes of a cold plasma with large chemical potential, which may characterize dense strongly-interacting matter in the core of neutron stars, was analysed in Ref.~\cite{Blaizot:1993}. For high-temperature plasmas, the hard-thermal-loop (HTL) approximation has been widely used in order to simplify the analysis of thermodynamics, transport, and collective behaviour of the QGP~\cite{Blaizot:2002}.  For non-thermal systems, one can use scale separation to define the so-called hard-loop (HL) approximation which relaxes the need for thermal equilibrium~\cite{Mrowczynski:2000ed, Romatschke:2003ms, Schenke:2006, Rebhan:2009}.

The effects of momentum anisotropy on in-medium quark and gluon propagation was studied in Ref.~\citep{Mrowczynski:2000ed} by calculating gluon and quark self-energies using both the HL diagrammatic method and the semi-classical kinetic transport theory approach in the case of gluons. The authors demonstrated that the two methods give equivalent results for the gluon polarization tensor. For the quark self-energy, they obtained a general result in the form of an integration over the 3-vector momentum of partonic degrees of freedom for anisotropic plasma. They did not further calculate the integrals necessary, however, they did specify the general structure of the dispersion relations. 

Based on analysis of the collective modes in a momentum-space anisotropic QGP, the occurrence of a color plasma instability, dubbed the Chromo-Weibel instability, was pointed out in Ref.~\cite{Mrowczynski:1994} and a more phenomenological treatment of unstable modes in heavy-ion collisions was investigated in Ref.~\cite{Randrup:2003}.   Following these works, in Ref.~\citep{Romatschke:2003ms} the authors used a tensor decomposition method to determine the gluonic collective modes of a spheroidally anisotropic QGP and the presence of unstable modes for both prolate and oblate anisotropies was demonstrated analytically and numerically. The spheroidal anisotropy studied therein was defined by a direction-dependent momentum-space rescaling (contraction/stretching) of an isotropic momentum distribution. The possible role of color-field instabilities in the fast thermalization of the strongly-interacting anisotropic plasma has been discussed by several researchers \cite{Rebhan:2004ur, Arnold:2005, Rebhan:2005re, Romatschke:2006wg, Bodeker:2007fw, Rebhan:2008uj, Berges:2009, Berges:2009-2, Rebhan:2010, Kurkela:2011ti, Kurkela:2011ub, Attems:2012js, Berges:2013eia}.  For a recent review of color plasma instabilities, see Ref.~\cite{Mrowczynski:2016etf}.

The quark self-energy and collective modes of a spheroidally anisotropic system were studied previously in Ref. \citep{Schenke:2006}.  The authors demonstrated that, in contrast to the gluonic modes, there are no unstable fermionic modes. The lack of unstable fermionic modes is expected on physical grounds, due to their inability to condense due to the Pauli exclusion principle. In addition, they demonstrated that, in the high temperature limit, the results for the non-equilibrium hard-loop self-energies obtained using the real-time formalism obey the Kubo-Martin-Schwinger relations appropriate for the equilibrium case. As a result, in order to calculate the photon production rates from an anisotropic plasma in the hard-loop limit, one can use the same production rate formula as in the isotropic equilibrium case, provided that one modifies the self-energy to account for the momentum-space anisotropy. 

The related calculation of photon production was presented in Ref.~\cite{Schenke:2007}, where the both hard-momentum contributions (from Compton scattering and pair annihilation) and the soft contributions (from HL real-time self-energy calculation) were considered. The isotropic Bose-Einstein and Fermi-Dirac distributions were transformed to the corresponding spheroidally anisotropic distributions, and it was shown that the IR divergences of the hard contributions cancel the UV divergences of the soft part. Their results demonstrated that the photon production rate depends on the polar angle (rapidity) in a non-trivial manner due to the local-rest-frame momentum-space anisotropy of the QGP.  In addition, the resulting angular dependence of the photon rates was found to increase with increasing photon energy.  In recent years, with the development of viscous anisotropic hydrodynamic models for the evolution of the QGP created in heavy-ion collisions \cite{Strickland:2014pga}, it is now possible to  self-consistently fold together the anisotropic rates for photon \cite{Bhattacharya:2016} and dilepton production \cite{Ryblewski:2015hea}.  These works are complementary to the approaches which use second-order viscous hydrodynamics \cite{Dion:2011, McLerran:2014, Shen:2015,  Paquet:2016} and parton/hadron transport \cite{Linnyk:2015tha, Linnyk:2015rco}.

One limitation of the prior works \cite{Bhattacharya:2016,Ryblewski:2015hea} is that they relied on a spheroidal approximation for the quark and gluon distribution functions.  While this is probably sufficient for understanding the effects of the rapid longitudinal expansion of the QGP on integrated electromagnetic rates, it may not be sufficient for understanding the elliptic flow of photons and dileptons which is on the order of a few percent.  The spheroidal parametrization of the momentum-space anisotropy, as initiated in Ref. \cite{Romatschke:2003ms}, has been used to address several aspects of anisotropic QGP \cite{Schenke:2007, Mandal:2012, Mandal:2014, Carrington:2014, Hong:2014,Thakur:2014, Carrington:2015, Krouppa:2015, Bhattacharya:2016}; however, the spheroidal parametrization (with one anisotropy direction) ignores the possibility of momentum anisotropies in the transverse directions. This may be particularly important in studying photon collective flow, as one needs to calculate the dependence of the quark self-energy and corresponding photon production rates on both the azimuthal and polar angles.  The generalization of the anisotropic distribution function has been considered in several prior works, see e.g. \cite{Tinti:2014, Nopoush:2014pfa, Tinti:2014yya}, however, in those works the authors focused on the hydrodynamics formalism itself and did not consider the quark or gluon self-energies.  In this paper, we calculate the quark self-energy by considering an ellipsoidally momentum-anisotropic distribution function with three perpendicular anisotropy directions.  This work sets the stage for a more self-consistent calculation of photon production and collective flow from an anisotropic QGP.

This paper is organized as follows: In Section \ref{quarkse}, the calculation of  quark self-energy in an ellipsoidally momentum-anisotropic QGP is presented.  The plots of the results for different angles and anisotropy strengths are shown and discussed in Section \ref{resultssec}. In section \ref{sec:conclusions}, we summarize the results and presents our concluding remarks and an outlook for future studies. Finally, in Appendix \ref{app:alternative} we present an alternative method for calculating the anisotropic quark self-energy which was used to cross-check our results contained in the main body of the text. In Appendix \ref{app:expansion}, we present the analytic formula for the quark self-energy in the limit of small anisotropies.  These expressions were used to cross check our numerical procedures in the small anisotropy limit.
 
\section{Anisotropic quark self-energy}
\label{quarkse}

The general expression for the gauge-independent retarded quark self-energy in a momentum-anisotropic system in the hard-loop (HL) approximation was first obtained in Ref. \cite{Mrowczynski:2000ed}
\ba 
 \Sigma(K) = \frac{C_F}{4} g^2 \int_{\bf p} \frac{f ({\bf p})}{|{\bf p}|} \frac{P \cdot \gamma}{P\cdot K} \, ,\label{retself}
\ea
where $P = (\omega_p,{\bf p})$ and $K = (\omega,{\bf k})$ are the Minkowski-space partonic momentum four-vectors, $C_F \equiv (N_c^2 -1)/2N_c$, $\int_{\bf p} \equiv \int d^3 p/(2 \pi)^3 $, $g$ is the QCD coupling, and the distribution function $f ({\bf p}) $ is the sum of the momentum distributions for quark and gluon partons $f ({\bf p}) \equiv 2 \left( n({\bf p}) + \bar n ({\bf p})
\right) + 4 n_g({\bf p})$.  

\subsection{Ellipsoidal self-energy setup}

Generalizing the setup used in Refs.~\cite{Romatschke:2003ms, Schenke:2006}, herein we require the local rest frame distribution function $f({\bf p})$ to be parametrized by
\be
f({\bf p})=f_{\boldsymbol\xi}({\bf p}) = f_{\rm iso}\left(\frac{1}{\lambda}\sqrt{{\bf p}^2+\xi_x({\bf p}\cdot{\bf \hat x})^2+\xi_y({\bf p}\cdot{\bf \hat y})^2+\xi_z({\bf p}\cdot{\bf \hat z})^2}\right)%
,
\label{squashing}
\ee
where $\hat{x}$, $\hat{y}$, and $\hat{z}$ are Cartesian unit vectors in the local rest frame of the matter, $\pmb{\xi}\equiv(\xi_x,\xi_y,\xi_z)$ are anisotropy parameters corresponding to three spatial dimensions, and $\lambda$ is a temperature-like scale. In this parametrization, $f_{\rm iso}$ is a general isotropic distribution function which reduces to the appropriate equilibrium distribution function in the isotropic equilibrium limit ($\pmb{\xi}=0$).  The anisotropy parameters $\xi_x$ and $\xi_y$ characterize the strength of anisotropy in transverse plane and $\xi_z$ characterizes the strength of anisotropy in the longitudinal direction. In other words, the spherical equal occupation number surfaces (isosurfaces) in momentum-space for the isotropic case transform to ellipsoidal isosurfaces in the anisotropic case. Using Eq.~(\ref{squashing}) one obtains
\ba \label{q-self2}
\Sigma(K) = \frac{m_q^2}{4\pi} \int d\Omega \, \Big(1+\xi_x({\hat p}\cdot{\bf \hat x})^2+\xi_y({\hat p}\cdot{\bf \hat y})^2+\xi_z({\hat p}\cdot{\bf \hat z})^2 \Big)^{-1}
\frac{P \cdot \gamma}{P\cdot K} \, ,
\ea
where
\ba
m_{\rm q}^2 = \frac{g^2 C_F}{8 \pi^2} \int_0^\infty dp \,
    p \, f_{\rm iso}\Big(\frac{p}{\lambda}\Big) \, .
\ea
As a result, all dependence on the form of the underlying isotropic distribution function is subsumed into the numerical value of $m_q$.

\subsection{Dirac decomposition and collective modes}

The self-energy (\ref{q-self2}) can be expanded as
\ba \Sigma(K) = \gamma^0 \Sigma_0 + {\boldsymbol\gamma}\cdot{\mathbf
\Sigma}\,,
\ea
where $\gamma^\mu$ are Dirac matrices. The quark collective modes are determined by finding all four-momenta $K$ for which the determinant of the inverse propagator $S$ vanishes
\ba
{\rm det}\,S^{-1} = 0 \; ,
\ea
where
\ba
i S^{-1}(K) &=& \gamma^\mu k_\mu - \Sigma(K) \, , \nonumber \\
&\equiv& \gamma^\mu \Delta_\mu \,,
\ea
with $\Delta(K)\equiv(\omega - \Sigma_0,{\bf k} - {\bf \Sigma})$.
Using the fact that ${\rm det}(\gamma^\mu \Delta_\mu) = (\Delta^\mu \Delta_\mu)^2$ and defining $\Delta_s^2 = {\bf \Delta}\cdot{\bf \Delta}$, the dispersion relations for the quark collective modes becomes 
\ba
\Delta_0 = \pm \Delta_s \, .
\label{fermiondisp}
\ea

\subsection{Calculation of the ellipsoidal quark self-energy}

We now turn to the explicit calculation of the self-energy (\ref{q-self2}) for an ellipsoidally anisotropic distribution function.   In the high-energy limit, to good approximation, one can ignore the quark bare masses and, as a result, the system is approximately conformal.  In the ellipsoidally-anisotropic case, for a conformal system there are only two physical anisotropy directions (transverse and longitudinal), and we can set $\hat{x}$ to be in the direction of transverse anisotropy and $\hat{z}$ to be the direction of longitudinal anisotropy. This amounts to rearranging the parameters and setting
\be
 \frac{\xi_z-\xi_y}{1+ \xi_y} \longrightarrow\ \xi_1\ , \ \ \ \frac{\xi_x -\xi_y}{1+\xi_y} \longrightarrow\ \xi_2\ , \ \ \ \frac{\lambda}{\sqrt{1+\xi_y}} \longrightarrow\ \lambda\,.
\label{xitransform} 
\ee
As a result, the quark self-energy can be written as
\be
\Sigma^{i} (K)= \frac{m_q^2}{4\pi} \int_0^{2\pi}d\phi\ \int_{-1}^{1}dx\ \frac{1}{A(1+s \cos^2\phi)}\ \frac{v^i}{a-b \cos\phi - c \sin\phi}\ , \\
\label{sigmaoverangles}
\ee
where
\ba
a&=&\frac{\omega}{k} - x \cos \theta_k\ , \\
b&=& \sin\theta_k \cos\phi_k \sqrt{1-x^2}\ , \\
c&=& \sin\theta_k \sin\phi_k \sqrt{1-x^2}\ , \\
A&=& 1+ \xi_1 x^2\ , \\
s&=& \frac{\xi_2 (1-x^2)}{A}\ , \\
v&=&(1,\ \sqrt{1-x^2} \cos\phi,\ \sqrt{1-x^2} \sin\phi,\  x )\ .
\label{eq:coeff}
\ea
Splitting the integrand of (\ref{sigmaoverangles}) using a partial-fraction decomposition, the self-energy components can be written as 
\be
\Sigma^{i}=\frac{m^2_q}{4\pi k} \int_{-1}^{1} dx\ \sum_{j=1}^{8}\ \lambda_{j}^{i}\ D_j \ \ \ \ \ \ \ (i=0,1,2,3) \, ,
\label{sigmasplit}
\ee
in which we have defined
\ba
D_1&=&\int_0^{2\pi}d\phi\ \frac{1}{a-b \cos\phi -c \sin\phi}\ , \\
D_2&=&\int_0^{2\pi}d\phi\ \frac{\cos\phi}{a-b \cos\phi -c \sin\phi}\ , 
\ea 
\ba 
D_3&=&\int_0^{2\pi}d\phi\ \frac{\sin\phi}{a-b \cos\phi -c \sin\phi}\ , \\
D_4&=&\int_0^{2\pi}d\phi\ \frac{\cos\phi \sin\phi}{a-b \cos\phi -c \sin\phi}\ , \\
D_5&=&\int_0^{2\pi}d\phi\  \frac{1}{1+ s \cos^2\phi}\ , \\
D_6&=&\int_0^{2\pi}d\phi\  \frac{\cos\phi}{1+ s \cos^2\phi}\ , \\
D_7&=&\int_0^{2\pi}d\phi\  \frac{\sin\phi}{1+ s \cos^2\phi}\ , \\
D_8&=&\int_0^{2\pi}d\phi\  \frac{\cos\phi \sin\phi}{1+ s \cos^2\phi}\ .
\label{Dintegrals}
\ea
One can see that for $s\geq -1$ (which is always the case, based on the limits of the anisotropy parameters), $D_6= D_7 = D_8 \equiv 0$. 
 The coefficients $\lambda_{j}^{i}:=({\boldsymbol \lambda}^i)_j $ can be obtained by solving the linear equation $L {\boldsymbol \lambda}^i =M^i$, where the elements of the 8-dimensional vector
$M^i$ are defined as $(M^i)_j=\delta^{i+1}_j $ ($\delta^{i}_j $ is Kronecker's delta), and the $8\times 8$ matrix $L$ is 
\begingroup
\[
\renewcommand\arraystretch{0.6}
L =
\begin{pmatrix}
1& 0& 0& 0& a& 0& -c& 0\\ 0& 1& 0& 0& -b& a& 0& -c\\ 0& 0& 1& 
    0& -c& 0& a& 0\\ 0& 0& 0& 1& 0& -c& -b& a\\ s& 0& 0& 0& 0& -b& 
    c& 0\\ 0& 0& s& 0& 0& 0& 0& -b\\ 0& s& 0& 0& 0& 0& 0& c\\ 0& 0&
     0& 1& 0& 0& 0& 0
\end{pmatrix}\,.
\]
\endgroup
Using this method, one can express the self-energy components as one-dimensional integrals over the variable $x\equiv\cos\theta$ 
\ba
\Sigma_0 &=&\frac{m^2_q}{2 k} \int_{-1}^{1} dx\  \frac{1}{A \rho} \Bigg[\alpha_0\sqrt{\frac{(a+b)^2}{a^2-(b^2+c^2)}}  + s \beta_0 \sqrt{\frac{1}{1+s}}\Bigg]\, , \\
\Sigma_x &=&\frac{m^2_q}{2 k} \int_{-1}^{1} dx\   \frac{1}{A \rho} \Bigg[\alpha_x\sqrt{\frac{(a+b)^2}{a^2-(b^2+c^2)}}  + s \beta_x \sqrt{\frac{1}{1+s}}\Bigg]\, , \\
\Sigma_y &=&\frac{m^2_q}{2 k} \int_{-1}^{1} dx\   \frac{1}{A \rho} \Bigg[\alpha_y\sqrt{\frac{(a+b)^2}{a^2-(b^2+c^2)}}  + s\beta_y \sqrt{\frac{1}{1+s}}\Bigg]\, , \\
\Sigma_z &=&\frac{m^2_q}{2 k} \int_{-1}^{1} dx\   \frac{1}{A \rho} \Bigg[\alpha_z\sqrt{\frac{(a+b)^2}{a^2-(b^2+c^2)}}  + s \beta_z \sqrt{\frac{1}{1+s}}\Bigg]\, ,
\ea
where we have defined
\ba
\rho&\equiv& (b^2+c^2)^2 +s (2 a^2 b^2 -2 a^2 c^2 + 2 b^2 c^2 + 2 c^4) + s^2 (a^2-c^2)^2\ , \\
\alpha_0&\equiv&\frac{1}{a+b}\Big[ (b^2+c^2)^2 + s(a^2 b^2 - a^2 c^2 +b^2 c^2 + c^4) \Big]\ , \\
\beta_0&\equiv& a(b^2-c^2) +s a(a^2 -c^2)\ , \\
\alpha_x &\equiv& \frac{ab}{a+b} \Big[(b^2+c^2)+s(a^2- c^2) \Big] \sqrt{1-x^2}\ , \\
\beta_x&\equiv&-b \Big[(b^2+c^2) + s(a^2+ c^2) \Big] \sqrt{1-x^2}\ , \\
\alpha_y&\equiv& \frac{ac}{a+b}\Big[(b^2+c^2) + s(-a^2 + 2b^2 +c^2) \Big] \sqrt{1-x^2} \ , \\
\beta_y&\equiv& -c \Big[(b^2+c^2) + s(-a^2+b^2+2 c^2) + s^2 (c^2-a^2) \Big] \sqrt{1-x^2} \ , \\ 
\alpha_z&\equiv& x \alpha_0\ , \\
\beta_z&\equiv& x \beta_0\ .
\ea

In the general case, we perform the integration over $x$ numerically for a given quark 4-momentum $K$. An alternative method for calculating the quark self-energy is presented in App.~\ref{app:alternative}. We also provide analytic results for the special case of small anisotropy parameters in App.~\ref{app:expansion}. 

\section{results}
\label{resultssec}

In this section, we present results for the components of the quark self-energy as a function of phase velocity $\omega/k$. In what follows, the real and imaginary parts of the four components of the quark self-energy are normalized by the quantity $m_{\rm q}^2/k$. Then, for presentation purposes, each individual component of the quark self-energy is scaled by a trivial geometrical factor which depends on the particular component being considered.   Following this scaling procedure, we consider the following quantities
\ba
\bar{\Sigma}_0&\equiv&\frac{k\Sigma_0}{m_{\rm q}^2}\,,\\
\bar{\Sigma}_x&\equiv&\frac{1}{\sin\theta_k\cos\phi_k}\frac{k\Sigma_x}{m_{\rm q}^2}\,,\\
\bar{\Sigma}_y&\equiv&\frac{1}{\sin\theta_k\sin\phi_k}\frac{k\Sigma_y}{m_{\rm q}^2}\,,\\
\bar{\Sigma}_z&\equiv&\frac{1}{\cos\theta_k}\frac{k\Sigma_z}{m_{\rm q}^2}\,.
\ea 
%
\begin{figure}[H]
\centerline{
\includegraphics[width=0.95\linewidth]{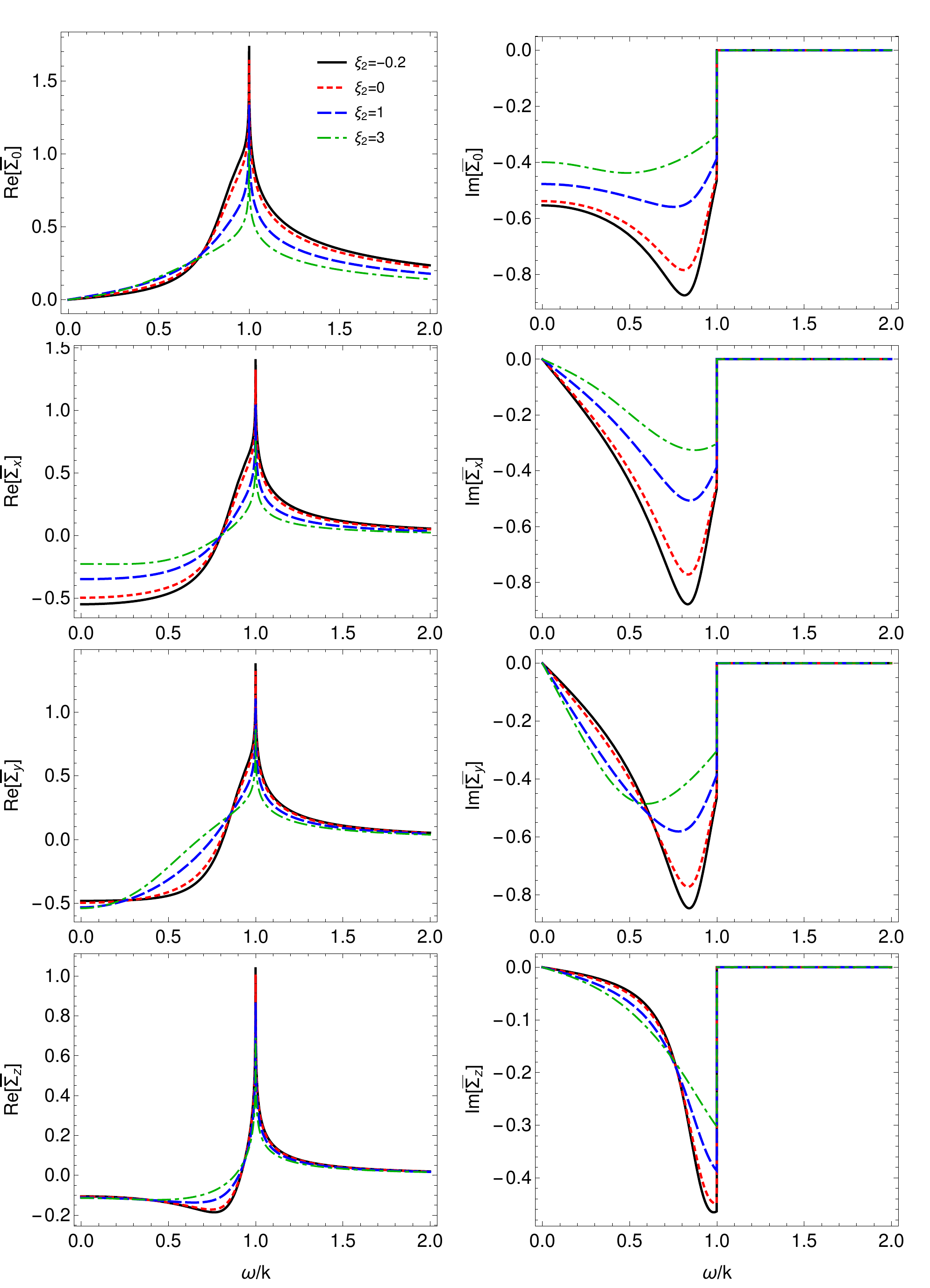}
}
\caption{The real and imaginary parts of $\bar{\Sigma}_0$, $\bar{\Sigma}_x$, $\bar{\Sigma}_y$, and $\bar{\Sigma}_z$ as a function of $\omega/k$ for $\xi_1=10$, $\theta_k=\pi/3$, $\phi_k=\pi/6$, and $\xi_2=\{-0.2,0,1,3\}$.} 
\label{plot:plot1}
\end{figure}

\begin{figure}[H]
\centerline{
\includegraphics[width=0.95\linewidth]{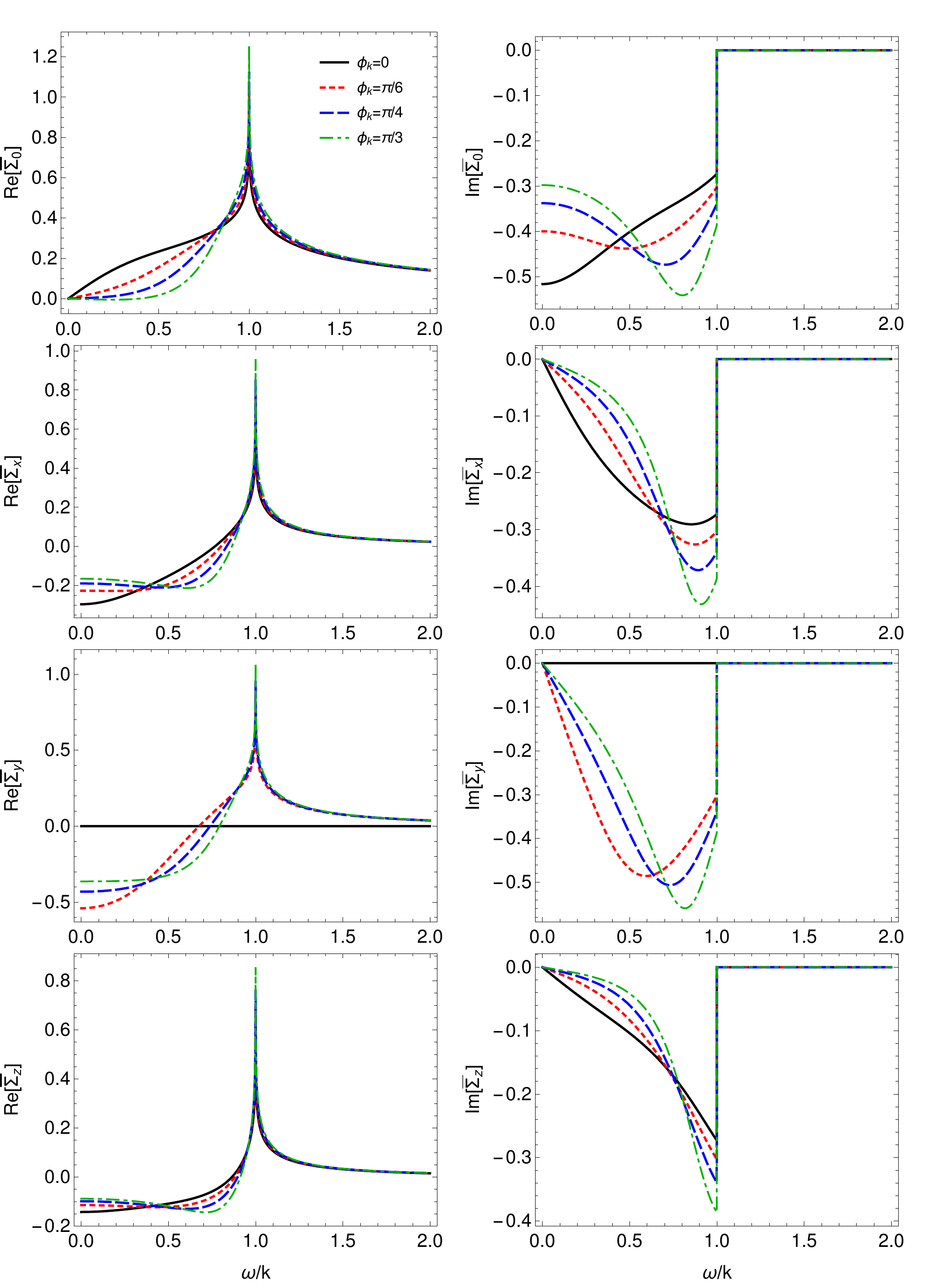}
}
\caption{The real and imaginary parts of $\bar{\Sigma}_0$, $\bar{\Sigma}_x$, $\bar{\Sigma}_y$, and $\bar{\Sigma}_z$ as a function of $\omega/k$ for $\xi_1=10$, $\xi_2=3$, $\theta_k=\pi/3$, and $\phi_k=\{0,\pi/6,\pi/4,\pi/3\}$.}
\label{plot:plot2}
\end{figure}

We have used the both the expressions presented in the previous section and those listed in App.~\ref{app:expansion} to obtain our numerical results; however, despite the fact that the expressions in App.~\ref{app:expansion} are more complicated, they are more straightforward to evaluate numerically since it is easier to identify the position of all poles in the integration domain.  Because of this, for the result plots presented in this section, we have used the method presented in App.~\ref{app:expansion}; however, in all cases we have verified explicitly that the two methods give the same results to within the required numerical precision.  Generally speaking, we find that the analytic structure of fermion self-energy is the same as in the anisotropic case, namely that for time-like momenta, $\omega/k > 1$,  the self-energy is real-valued and for space-like momenta, $\omega/k<1$, there is a cut in the complex plane which spans the line $\Im[\omega/k]=0$.

In Fig.~\ref{plot:plot1}, we present the components of the scaled quark self-energy for $\xi_1=10$, $\theta_k=\pi/3$, $\phi_k=\pi/6$, while varying the transverse anisotropy parameter with $\xi_2=\{-0.2,0,1,3\}$. As can be seen in this plot, the real part of the components of the quark self-energy tend to zero for large $\omega/k$, while the imaginary parts drop to zero abruptly for $\omega/k > 1$ due to the absence of the Landau cut for time-like momenta. The plots also show that the magnitude of self-energy components depend on the magnitude of the transverse anisotropy, as one can expect on general grounds.  This dependence would be reflected in a photon production rate that possesses explicit azimuthal anisotropies which are independent from those generated solely due to QGP collective flow.  To demonstrate this feature more explicitly, in Fig.~\ref{plot:plot2} we use $\xi_1=10$, $\xi_2=3$, $\theta_k=\pi/3$ and vary the azimuthal scattering angle as $\phi_k=\{0,\pi/6,\pi/4,\pi/3\}$. 

\section{Conclusions and Outlook}
\label{sec:conclusions}

In this paper, we determined the self-energy of quarks in an ellipsoidally-anisotropic QGP by using the method of partial-fraction decomposition together with numerical evaluation of the resulting one-dimensional integrals. Previous results for the hard-loop self-energy of quarks in a spheroidally-anisotropic QGP were extended by generalizing the parametrization of the momentum distribution functions to incorporate anisotropies in transverse momentum-space directions. 

 With the introduction of the additional anisotropies in the transverse plane, the calculations become a bit more tedious compared to the case of a spheroidal momentum anisotropy, however, the final results can be expressed as modifications of the previously considered case.  Our results show that anisotropies in transverse momentum directions affect the quark self-energy, as can be expected on general grounds and herein we demonstrated how to evaluate the effects quantitatively.  We have shown that the self-energy modifications due to transverse anisotropies induce additional angular dependence of the self-energy in transverse-momentum plane. As a result, there might be observable effects of an ellipsoidal momentum anisotropy in heavy-ion collision experiments.   In particular, the transverse anisotropies can introduce azimuthal angular dependence in the photon production rate, which would result in explicit azimuthal anisotropies in photon production, e.g. elliptic flow, triangle flow, etc.  This source of azimuthal anisotropy is distinct from that induced solely by the collective flow of the QGP itself and is, instead, directly related to viscous effects.

As a demonstration of the underlying source of the effect, we presented the variation of both the real and imaginary parts of the quark self-energy for different combinations of the anisotropy parameters and azimuthal angles. Comparing to previous results obtained in the spheroidal case, in an ellipsoidally anisotropic system one observes modifications to the real part of self-energy which are related to the effective mass of quasi-particles.  As a result, quarks obtain effective masses which depend on their full 3d direction of propagation. We found that the effect on the imaginary part of self-energy, which is related to the decay or production rates of particles, is larger than the effect on the real part. These modifications will affect QGP differential photon production rates.

Looking to the future, the results obtained herein form the basis of a self-consistent calculation of photon production from a QGP as created in relativistic heavy-ion collisions.  The underlying anisotropic formalism guarantees that the photon production rate is positive-definite at all momenta, which is not guaranteed using typical viscous hydrodynamics approaches.  Anisotropic hydrodynamics codes which take into account ellipsoidal anisotropies already exist and the output of the space-time evolution of the momentum-space anisotropies $\pmb{\xi}$, hard-momentum scale $\lambda$, and the collective flow generated during QGP evolution can now be folded together to obtain the final photon spectra including the effect of explicit azimuthal anisotropies in the rate.  This will extend previous works \cite{Schenke:2007,Bhattacharya:2016} which employed a spheroidal approximation.  We leave the computation of the integrated photon spectra to future work.  Finally, we also note that the method of partial-fraction decomposition presented in this paper can also be applied to the gluon polarization tensor in an ellipsoidally anisotropic QGP.

\acknowledgments{M. Strickland and M. Nopoush were supported by the U.S. Department of Energy, Office of Science, Office of Nuclear Physics under Award No.~DE-SC0013470.}

\appendix 

\section{Alternative Derivation}
\label{app:alternative}

In this appendix we present an alternative method for deriving the quark self-energy subject to an ellipsoidal momentum-space anisotropy. In comparison to the method presented in the body of the text, this method is based on three anisotropy parameters corresponding to two transverse and one longitudinal directions. Expanding the relation (\ref{q-self2}), one finds the following relation
\ba 
\Sigma^i(K) = \frac{m_q^2}{4\pi k}\!\int_{-1}^{1}\!dx\!\int_0^{2\pi}\!\frac{d\phi}{c_x \cos^2\phi+c_y \sin^2\phi+c_z} \frac{v^i}{a-b\cos\phi-c\sin\phi} \,,
\ea
with a,b, and c defined in Eq.~(\ref{eq:coeff}) and the variables $c_x$, $c_y$, and $c_z$, which are independent of $\phi$, being defined as
\ba
c_x&\equiv& \xi_x(1-x^2)\, , \\
c_y&\equiv& \xi_y(1-x^2)\, , \\
c_z&\equiv& 1+\xi_z x^2\, .
\ea
Using partial-fraction decomposition, one can transform the integral over $\phi$ for each component of $\Sigma$ into four non-trivial simpler ones:
\ba
\Sigma^i=\frac{m_{\rm q}^2}{k}\sum_{j=1}^4 \int_{-1}^1 dx\,n_{j i}(x) {\cal I}_j(x) \ ; \ \ \ \ \ \ \ (i=t,x,y,z)\,.
\label{eq:sigma}
\ea
The ${\cal I}$-functions used here are defined as
\ba
{\cal I}_1(x)&\equiv&\frac{2}{a+r}\sqrt{\frac{a+r}{a-r}}\, ,\\
{\cal I}_2(x)&\equiv&1-\frac{a}{2}{\cal I}_1(x)\, , \\
{\cal I}_3(x)&\equiv&\frac{1}{\sqrt{c_2^2-c_1^2}}\, , \\
{\cal I}_4(x)&\equiv&-c_2 {\cal I}_3(x)+1\, ,
\ea
and 
\ba
r &\equiv&\sqrt{1-x^2}\sin\theta_k\,, \\
c_1&\equiv&c_x-c_y=\xi_a (1-x^2)\,,\\
c_2 &\equiv&c_x+c_y+2c_z=-\xi_b x^2+\xi_x+\xi_y+2\, ,
\ea
with $\xi_a\equiv \xi_x-\xi_y$ and $\xi_b\equiv \xi_x+\xi_y-2\xi_z$.
By defining $e\equiv -re^{i\phi_k}/2$, and $f\equiv -re^{-i\phi_k}/2$, and the following functions,
\ba
{\cal D} &\equiv& a^4 c_1^2 +{\cal R}^2(c_2,c_1)-2a^2c_1{\cal R}(c_1,c_2)\,,\\
{\cal R}(x_1,x_2) &\equiv& 2 x_1 e f -x_2 (e^2+f^2)\, .
\ea
The coefficients $n_{ji}$ used in Eq.~(\ref{eq:sigma}) are defined as
\ba
n_{1t}&=&\frac{e}{{\cal D}}\Big[a^2 c_1 e+f {\cal R}(c_2,c_1)\Big]=n_{1z}/x\, ,\\
n_{2t}&=&\frac{a c_1}{{\cal D}}\Big[e^2-f^2\Big]=n_{2z}/x\, ,\\
n_{3t}&=&\frac{a c_1}{{\cal D}}\Big[a^2c_1 +2f(-c_1 e +c_2 f)\Big]=n_{3z}/x\,, \\
n_{4t}&=&-n_{2t}=n_{4z}/x\,, \\
n_{1x}&=&-\frac{a e}{2{\cal D}}\Big[a^2 c_1+(c_1-c_2)(e^2-f^2)-{\cal R}(c_1,c_2)\Big]\, ,\\
n_{2x}&=&-\frac{(e-f)}{2{\cal D}}\Big[a^2 c_1-{\cal R}(c_2,c_1)\Big]\, ,\\
n_{3x}&=&\frac{1}{2{\cal D}}\bigg[a^2 c_1\Big(2c_2 e -c_1(e+f)\Big)+\Big(2c_2 f-c_1(e+f)\Big){\cal R}(c_2,c_1)\bigg]\,, \\
n_{4x}&=&-n_{2x}\,, \\
n_{1y}&=&-\frac{ia e}{2{\cal D}}\Big[a^2 c_1-(c_1+c_2)(e^2-f^2)-{\cal R}(c_1,c_2)\Big]\,, \\
n_{2y}&=&\frac{-i(e+f)}{2{\cal D}}\Big[a^2 c_1+{\cal R}(c_2,c_1)\Big]\, ,\\
n_{3y}&=&\frac{i}{2{\cal D}}\bigg[a^2 c_1\Big(2c_2 e +c_1(e-f)\Big)+\Big(2c_2 f-c_1(e-f)\Big){\cal R}(c_2,c_1)\bigg]\, ,\\
n_{4y}&=&-n_{2y}\,.
\ea

\section{Small anisotropy expansion}
\label{app:expansion}

For small anisotropy, one can Taylor-expand the quark self-energy around $\bm{\xi}=0$. In this limit, the integral can be calculated analytically. To leading order in the anisotropy parameters, one finds
\ba
\Sigma_0=\Sigma^{\rm iso}_0+\Big[{\cal F}_{0,1}+\Sigma^{\rm iso}_0{\cal F}_{0,2}\Big]\, , \\
\frac{\Sigma_x}{\sin\theta_k\cos\phi_k}=\Sigma^{\rm iso}_s+\Big[{\cal F}_{x,1}+\Sigma^{\rm iso}_s{\cal F}_{x,2}\Big]\, , \\
\frac{\Sigma_y}{\sin\theta_k\sin\phi_k}=\Sigma^{\rm iso}_s+\Big[{\cal F}_{y,1}+\Sigma^{\rm iso}_s{\cal F}_{y,2}\Big]\, , \\
\frac{\Sigma_z}{\cos\theta_k}=\Sigma^{\rm iso}_s+\Big[{\cal F}_{z,1}+\Sigma^{\rm iso}_s{\cal F}_{z,2}\Big]\, ,
\ea
where
\ba
\Sigma^{\rm iso}_0&=&\frac{m_{\rm q}^2}{2k} \log\frac{\omega+k}{\omega-k}\, , \\
\Sigma^{\rm iso}_s&=&\frac{m_{\rm q}^2}{k} \Big(\frac{\omega}{2k}\log\frac{\omega+k}{\omega-k}-1\Big)\,. 
\ea
The various functions are
\ba
{\cal F}_{0,1}&=&\frac{zm_{\rm q}^2}{8k}\Big[6\xi_a\cos2\phi_k\sin^2\theta_k-\xi_b(3\cos2\theta_k+1)\Big]\, ,\\
{\cal F}_{0,2}&=& \frac{1}{8}\Big[2\xi_a \cos2\phi_k\sin^2\theta_k-\xi_b\big(\cos2\theta_k+3\big)-8\xi_z\Big] -\frac{zk}{m_{\rm q}^2} {\cal F}_{0,1}\, , \\
{\cal F}_{x,1}&=&\frac{m_{\rm q}^2}{24k}\Big[\xi_a\big(10\cos2\phi_k\sin^2\theta_k-4\big)-\xi_b\big(5\cos2\theta_k+3\big)\Big]\,,\\
{\cal F}_{x,2}&=&\frac{1}{8}\Big[6\xi_a\cos2\phi_k\sin^2\theta_k-\xi_b(3\cos2\theta_k+1)-8\xi_x\Big]-3\frac{k z^2}{m_{\rm q}^2} {\cal F}_{x,1}\,,\\
{\cal F}_{y,1}&=&\frac{m_{\rm q}^2}{24k}\Big[\xi_a\big(10\cos2\phi_k\sin^2\theta_k+4\big)-\xi_b\big(5\cos2\theta_k+3\big)\Big]\,,\\
{\cal F}_{y,2}&=&\frac{1}{8}\Big[6\xi_a\cos2\phi_k\sin^2\theta_k-\xi_b(3\cos2\theta_k+1)-8\xi_y\Big]-3\frac{k z^2}{m_{\rm q}^2} {\cal F}_{y,1}
\,,\\
{\cal F}_{z,1}&=&\frac{m_{\rm q}^2}{24k}\Big[10\xi_a\cos2\phi_k\sin^2\theta_k-\xi_b(5\cos2\theta_k-1)\Big]\,,\\
{\cal F}_{z,2}&=&\frac{1}{8}\Big[6\xi_a\cos2\phi_k\sin^2\theta_k-\xi_b(3\cos2\theta_k+1)-8\xi_z\Big] -3\frac{k z^2}{m_{\rm q}^2} {\cal F}_{z,1}\,.
\ea
%
\bibliography{fermion}

\end{document}